\documentclass{eptcsstyle/eptcs}
\usepackage[appendix=strip]{apxproof}

\usepackage{amsmath}
\usepackage{hyperref}
\usepackage{longtable}
\usepackage{xcolor}
\usepackage{xspace}
\usepackage{stmaryrd}
\usepackage{todonotes}%
\usepackage[normalem]{ulem}
\usepackage{macros}

\usepackage{algpseudocode}
\usepackage{algorithm}
\usepackage{algorithmicx}
\usepackage{tikz}
\usetikzlibrary{arrows.meta,graphs}
\usetikzlibrary{arrows}
\usepackage{color}

\newtheoremrep{theorem}{Theorem}
\newtheoremrep{lemma}[theorem]{Lemma}
\newtheoremrep{proposition}[theorem]{Proposition}
\newtheoremrep{corollary}[theorem]{Corollary}
\newtheoremrep{definition}[theorem]{Definition}
\newtheoremrep{claim}[theorem]{Claim}
\newtheoremrep{example}{Example}

\usepackage{xparse}

\usepackage{booktabs} 

\begin{document}

\title{Jumping Evaluation of Nested Regular Path Queries}
\def\titlerunning{Jumping Evaluation of Nested Regular Path Queries}
\def\authorrunning{J. Niehren, S. Salvati \& R. Azimov}

\author{Joachim Niehren, Sylvain Salvati
     \institute{Inria Lille, Université de Lille, France}
     \and
      Rustam Azimov
  \institute{
    Saint Petersburg State University, Russia}
  }

  \def\authorrunning{J. Niehren, S. Salvati, R. Azimov}
\maketitle

\begin{abstract}

Nested regular path queries are used for querying graph databases
and \RDF triple stores. We propose a new 
algorithm for evaluating nested regular path queries
on a graph from a set of start nodes in
combined linear time. We show that this complexity upper bound
can be reduced by making it dependent on the size of the query's {\em
top-down needed\/} subgraph, a notion that we introduce. For many
queries in practice, the top-down needed subgraph is way 
smaller than the whole graph.
Our algorithm is based on a novel compilation schema from nested
regular path queries to monadic datalog queries. Its complexity
upper bound follows from known properties of top-down datalog evaluation.
As an application, we show that our algorithm permits to reformulate
in simple terms a variant of a very efficient automata-based algorithm
proposed by Maneth and Nguyen that evaluates navigational path queries
in datatrees based on indexes and jumping.
Moreover, it overcomes some limitations of Maneth and Nguyen's: it is not
bound to trees and applies to graphs; it is not limited to forward 
navigational XPath but can treat any nested regular path query and 
it can be implemented efficiently without any dedicated
techniques, by using any efficient top-down datalog evaluator.
We confirm the efficiency of our algorithm experimentally
based on an implementation with LogicBlox.

\end{abstract}




\renewcommand\subsection[1]{\paragraph{#1.}}

\section{Introduction}

Regular path queries \cite{martens_et_al:LIPIcs:2018:8594} 
are regular expressions for navigating in edge labeled graphs. They
belong to the core of various query languages for
graph databases  and \RDF triple stores. Nested
regular path queries (\NRPQs) \cite{10.1145/2448496.2448513}
extend on regular expressions by adding filters\ignore{What is exactly a filter? and, what is their evaluation?} with 
logical operators, that may again contain regular path 
queries. \NRPQs were first invented as the programs of 
propositional dynamic logic (\PDL)
\cite{DBLP:journals/jcss/FischerL79}. 
\NRPQs are also part of \nSPARQL for querying
knowledge stores in the Semantic Web~\cite{DBLP:journals/ws/PerezAG10}.
The restriction of \NRPQs to data trees constitutes the navigational core of
regular \XPath. 
The \NRPQ ${\edge_a}^*/\neg[{\edge_b}^*/\edge_c]?$, for instance,
selects all nodes of an edge-labeled graph, which can be reached from a given set of
start nodes over a sequence of $a$-labeled edges,
under the condition expressed by the filter
$\neg[{\edge_b}^*/\edge_c]$: there must not exist any outgoing sequence of edges satisfying the regular
expression ${\edge_b}^*/edge_c$.

The set of nodes that can be reached by an \NRPQ
$P$  on a graph $G$ with a set of start nodes $S$ can be computed in
combined linear time, i.e. in linear time in the
  product of the inputs $\mathcal{O}(|P||G|)$.
  This complexity upper bound is folklore
  in the context of database theory, \XPath, and \nSPARQL,
  and was known already for \PDL,  but 
was first shown for the richer alternation-free modal
$\mu$-calculus 
\cite{cleaveland91:_linear_time_model_check_algor}.
However, it is also satisfied by many inefficient
algorithms: only the relevant fraction of the graph database should be
  visited for answering a database
  query. Any efficient query answer algorithm 
  must avoid complete traversals of large graphs whenever possible.

  Which part of a graph is relevant for a answering a
query may depend on the query answering algorithm though.
Therefore, we formalize a notion of needed
subgraph coined as \emph{top-down needed subgraph}, as the subgraph
that is traversed with a top-down evaluation of the query.
We then search for a query answering algorithm with combined
linear complexity \emph{with respect to the top-down needed subgraph},
instead of the whole graph which we consider as too expensive.

For regular path queries (without filters) a canonical notion
of the top-down needed subgraph seems quite intuitive, since
a regular expression can be interpreted in a top-down manner
as a navigation plan for traversing
a graph. The top-down needed subgraph contains all the nodes and
edges that are traversed when executing this navigation plan on the
graph, while starting with the given set of start nodes.
This notion of top-down needed 
nodes can  then be lifted from to \NRPQs, so that
any filter of a \NRPQ is tested only for those nodes where it is
required when executing the query in a top-down manner.

For evaluating regular path queries (without filters) with the above
complexity requirement, it is sufficient to execute it top-down navigation
plan on the graph. We note that the presence
of the Kleene star makes memoization mandatory, otherwise 
the algorithm may loop infinitely. What becomes more tedious is to find an evaluation algorithm for \NRPQs
that satisfies our complexity requirement. The existing proposals  in \cite{DBLP:journals/ws/PerezAG10,arenas2011querying,GottlobKochPichler03c} 
achieve combined linear time complexity by pre-evaluating the filters all over the
graph in a bottom-up manner and then running  an evaluation algorithm for regular path queries. However, the
bottom-up pre-computation of the filters over all the graph may visit
nodes that are \emph{not} needed for the top-down evaluation of \NRPQs, 
so these algorithms do not satisfy the envisaged complexity bound.
%

\begin{figure}
  \begin{minipage}{\textwidth}
\begin{minipage}{.59\textwidth}
	\begin{center}
		\includegraphics[width=5cm]{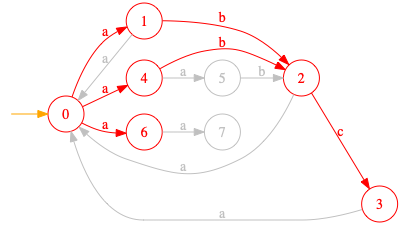} \vspace{-.6cm}
	\end{center}
    \caption{\label{graph}The graph $G_0$, the start set $S_0=\{0\}$,  and the top-down needed subgraph for
      $P_0 =\edge_a/[\edge_b/\edge_c]?$ in red.}
  \end{minipage}
  \quad
  \begin{minipage}{.35\textwidth}
      			$
			\begin{array}{l}
			q_{0}(x) \DL  q_{1}(x), q_{2}(x).\\
			q_{0}(x) \DL \start(y),\edge_a(y,x).\\
			q_{2}(x) \DL \edge_b(x,y), q_{3}(y).\\
			q_{3}(y) \DL \edge_c(y,z).\\
			\end{array} 
			$
                        \ignore{
			$
			\begin{array}{l}
			q_{0}(x) \DL  q_{1}(x), q_{2}(x).\\
			q_{1}(y) \DL \start(x),\edge_a(x,y).\\
			q_{2}(x) \DL \edge_b(x,y), q_{3}(y).\\
			q_{3}(y) \DL \edge_c(y,z).\\
			\end{array} 
			$
			}
			\caption{\label{datalog-for-P} The Datalog program $M_0$ for the nested
				regular path query 
				$P_0 $ from \Figure{graph}. \qquad}
\end{minipage}
  \end{minipage}
\end{figure}

As an example, we consider in \Figure{graph} the graph $G_0$ 
with edge labels $\{a,b,c\}$, the \NRPQ
$P_0=\edge_a/[\edge_b/\edge_c]?$, and the set of
start nodes $S_0=\{0\}$. The query $P_0$ started at $S_0$ selects all those
nodes of $G_0$ that are connected to the start node $0$ by an $a$-edge, 
and have a path over a $b$-edge followed by a $c$-edge.
The top-down algorithm with pre-evaluation of filters for $P_0$ will 
first compute the answer set of the filter $[\edge_b/\edge_c]$ on
$G_0$ starting with $S_0$, which is $\{1,4,5\}$. 
It will then compute the set of nodes that are reached from the 
start node $0$ over an $a$-edge which is $\{1,4,6\}$. 
The answer set is the intersection which is
$\{1,4\}$. This algorithm, however,  will inspect 
some nodes and edges for the
pre-evaluation of the filters  that are \emph{not} top-down needed, namely
the node $5$ and the $b$-edge from $5$ to $2$.
So the difficulty is to avoid the bottom-up pre-evaluation
of filters.

We will show that \NRPQs can be evaluated with the expected complexity
by enhancing the naive top-down evaluator for \NRPQs with memoization -- 
instead of precomputing the filters. We obtain the right
kind of memoization by compiling the path query 
into a monadic datalog program, and then evaluating this datalog 
program in a top-down manner with memoization. We note that
a monadic datalog program may still use extensional predicates of
higher arities, like the predicates $\edge_a$, $\edge_b$, and
$\edge_c$ in our example. While the existence
of compilers from \NRPQs to monadic datalog is less
surprising -- even though none  was published earlier to the best of
our knowledge -- the main difficulty is to find a datalog program
that captures the top-down neededness with respect to the \NRPQ.
In the case of $P_0$ we obtain the datalog program  in
\Figure{datalog-for-P}.  We suppose that the table of the
monadic extensional predicate $\start$ contains the
set of start nodes in $S_0$.
We note that paths in filters such as $[\edge_b/edge_c]$ are
compiled quite differently to paths outside filters. The reason is that
paths outside filter have to return the end node that is reached,
while paths inside filters have to jump back from the end node
to the start node. The binary relation between end nodes and
start nodes, however, cannot be stored in any predicate of the
datalog program, since
this would require a binary intensional predicate that are
ruled out by monadic datalog. What we exploit
instead is that a monadic datalog program can perform multiple
tests on the same node, as with the rule $q_{0}(x) \DL  q_{1}(x), q_{2}(x).$
In our example, $q_1(x)$ will bind $x$ to some
node reached  over the path $\edge_a$ from the set of start nodes
$S_0$, while $q_2(x)$ will test the filter
$[\edge_b/edge_c]$ there.

Our first contribution is an algorithm that answers \NRPQs in the time
$\mathcal{O}(|\tdnGS(P)||P|)$ where $|\tdnGS(P)|$ is the size of
the top-down needed subgraph of $G$ for query $P$. For this, we
present a novel linear time compilation scheme  mapping path 
queries to datalog queries.  
We restrict ourselves to negation-free \NRPQs in order to avoid the
usage of stratified negation for the sake of simplicity.
We prove that if the compiler transforms a query
$P$ and a start set $S$ into a datalog query $M$, then the top-down needed 
subgraph $\tdnGS(P)$ is the part of the graph's database that is
visited by top-down evaluation of the datalog query $M$ on the database.  
Furthermore, the datalog queries produced are monadic and
restricted in such a way, that their top-down evaluation can be 
done in combined linear time depending on the size of the top-down 
visited  subdatabase. 
It follows 
that the answer set of an \NRPQ $P$ on a graph $G$ with start set
$S$ can indeed be computed in time $\mathcal{O}(|\tdnGS(P)| |P|)$.

Our algorithm can be extended to a jumping algorithm for
answering \NRPQs on graphs with indexes. The indexes  are 
binary relations defined by other \NRPQs that allow the algorithm
to jump in the graph. For instance,
when given an index for the \NRPQ $I = \edge^*/\filtertopath{a}$ on
the input graph, the evaluation algorithm can always jump to all 
$a$-labeled nodes accessible from the current node, without visiting 
the intermediates.  We consider that the indexes are given 
with the input, since they are  usually pre-computed elsewhere. 
Therefore, the indexes can simply be integrated into the graph
as new edges that are labeled by the index's name, which is $I$ in our example.
Furthermore, the \NRPQ is then rewritten by substituting all occurrences of
$I$ as a subquery in the \NRPQ by $\edge_I$, so that we can 
apply the previous machinery.
An efficient implementation of our algorithm can be
based on any efficient top-down datalog evaluator, since it is sufficient
to evaluate the monadic datalog program produced by
our compiler. 

Our graph jumping algorithm permits
 to reformulate without specialized techniques a very 
efficient  automata-based algorithm proposed by 
\cite{ManethNguyen10} that evaluates \NRPQs on
datatrees with indexes based on jumping. More precisely, their algorithm covers forward navigational \XPath
queries on \XML documents. It is based on alternating tree automata
with selection states (which can be seen as binary datalog programs
while ours are monadic). Our generic approach overcomes the
limitations of their algorithm: it is not bound to trees but applies to graphs;
it is not limited to forward 
navigational XPath but can treat any \NRPQs also with backward steps.
it can be implemented efficiently without any dedicated 
techniques, by using any efficient datalog evaluator supporting
top-down evaluation such as LogicBlox \cite{LogicBlox-short}.

\ignore{Furthermore, any efficient semi-naive bottom-up
evaluator will do the 
job, since we can rely on the magic set transformation
to reduce top-down evaluation to bottom-up evaluation.
The point here is that the magic set transformation
of the datalog programs produced by our compilation schema
can be done in linear time since these programs are monadic.
For positive path queries, where the 
resulting datalog program is negation free, we can
apply Theorem 3 of \cite{Ullman89},
showing that bottom-up evaluation after the magic set 
transform is more efficient than top-down evaluation.
In the case with negation, we can use a recent generalization
of this theorem to stratified datalog
programs~\cite{tekle19:_exten_magic_negat}.} 

\noindent\textit{\textbf{Outline.}} In Section
\ref{sec:NRPQs}, we recall the definition of \NRPQs.
In Section \ref{sec:tdn}, we formally define top-down needed subgraphs. In Section \ref{sec:datalog}, we recall preliminaries
on datalog queries, while discussing the complexity
of top-down evaluation in Section \ref{sec:complexity}.
In Section \ref{sec:compiler}, we give our compiler from
\NRPQs to datalog queries with its complexity theorem.
Proofs can be found in the appendix.
Section~\ref{sec:jumping} presents the jumping 
evaluation algorithm for \NRPQs on graphs
with indexes, and Section~\ref{sec:expe}
preliminary experimental results.

\section{Nested Regular Path Queries}
\label{sec:NRPQs}

\label{sec:syntax-semantics}
Regular path queries on labeled graphs \cite{10.1145/2448496.2448513} 
can be extended to \NRPQs by adding filters with  
logical operators \cite{martens_et_al:LIPIcs:2018:8594}.
CoreXPath \cite{GottlobKochPichler03c}
is a sublanguage of \NRPQs with limited recursion
where the interpretation is restricted to an unranked tree.
\NRPQs were known even much earlier as the propositional
dynamic logic (\PDL) of \cite{DBLP:journals/jcss/FischerL79}.

We start from a finite set of labels $\Sigma$. A (finite)
$\Sigma$-labeled digraph is a tuple $G=(V,(V_a)_{a\in \Sigma},$
$(E_a)_{a\in\Sigma})$ where $V$ is a finite set of nodes,
$V_a\subseteq V$ a finite subset of $a$-labeled nodes,
and $E_a \subseteq V\times V$  a finite set of
$a$-labeled edges where $a\in\Sigma$. Note that nodes may have multiple labels
or none, while each edge has a unique label. Between
two nodes there may be multiple edges with different
labels though. 
An example for a labeled graph $G_0$ with labels in $\Sigma=\{a,b,c\}$
was given graphically in \Figure{graph}. The set of nodes of the graph
is $V=\{0, \ldots,7\}$. Here, the nodes are not labeled, so
$V_a=V_b=V_c=\emptyset$. Each of the edge has
a unique label. There are 8 $a$-labeled edges in $E_a$,
3 $b$-labeled edges in $E_b$ and one $c$-labeled
edge in $E_c$.

The syntax of \NRPQs with labels in $\Sigma$ is presented in
\Figure{syntax}. It consists
of a set of  filters $\FI$ that select a set of graph nodes, and a set of paths $\PA$
that select a set of pairs of graph nodes.
\begin{figure*}[t]
	$
	\begin{array}{lcl}
	\textit{filters} \qquad F \in \FI &::=& [P] \hmid \node \hmid \nodea\hmid F \land F' 
	\hmid F\vee F'\hmid \neg F\\                          
	\textit{paths} \qquad P \in \PA &::=& \filtertopath{F} \hmid \edge_a\hmid
	\edge_a\I \hmid
	P / P' \hmid P\cup P' \hmid
	P^+ \hmid goto(F) 
	\end{array}
	$
	\caption{\label{syntax} The syntax of \NRPQs with labels $a \in \Sigma$.}
\end{figure*}
The filter $\node$ selects all the nodes, while the filter $\nodea$
selects all $a$-labeled nodes. The set of nodes that are both $a$-labeled and $b$-labeled but 
not $c$-labeled is queried by 
filter $\nodea\wedge\node_b\wedge\neg\node_c$.
Path $\edge_a$ selects all $a$-labeled edges and
path $\edge=_\DEF\cup_{a\in\Sigma}\edge_a$ the set of
all edges. The path $\filtertopath{\True}$ selects
the identify on nodes $\{(v,v)\mid v\in V\}$.
Path composition $P/P'$, path union $P\cup P'$
are supported as well as repeated path composition $P^+$.
The Kleene star on paths can be defined by $P^*=_\DEF
P^+\cup \filtertopath{\True}$. Backwards edges
can be queried by $\edge_a\I$, so that general 
backwards path $P\I$ can be defined, where
$(P_1/P_2)\I= P_2\I/P_1\I$ and $\filtertopath{F}\I=\filtertopath{F}$. 
Finally, the path $goto(F)$ permits to jump to any node 
of the graph satisfying filter $F$. In particular, if
there is a label $root\in\Sigma$ that distinguishes
a set of roots, than path $goto(\node_{root})/P$ first
jumps to some root node before executing path
$P$.

A little more complex example for an \NRPQ  with signature $\Sigma=\{a,b,c\}$
is the path query
$P_2 = \filtertopath{\nodea}/(\edge^+/\filtertopath{[\edge_b/\edge_c]})^*.$
The evaluation of $P_2$ on a given graph from a
start node tests whether the start 
node is $a$-labeled, and if so, it navigates from
there repeatedly, over a sequence of edges to some 
node for which there exists an outgoing path over edges
with labels $b$ and then $c$.
The set of all nodes reached this way
is selected.

\begin{figure*}[t]
	\begin{minipage}{\textwidth}
		\begin{minipage}{.5\textwidth}
			$
			\begin{array}{l} 
			\Sem{[P]}  = \{v\mid \exists v'.\ (v,v')\in \Sem{P} \}\\
			\Sem{\node}  = V\\
			\Sem{\nodea}  = V_a\\
			\Sem{\neg F}  = V\setminus \Sem{F} \\
			\Sem{F\wedge F'}  = \Sem{F}\cap \Sem{F'} \\
			\Sem{F\vee F'}  = \Sem{F}\cup \Sem{F'} \\
			\end{array}
			$
		\end{minipage}
		\begin{minipage}{.49\textwidth}
			$
			\begin{array}{l}	
			\Sem{\filtertopath{F}} = \{ (v,v) \mid v \in \Sem{F}\}\\
			\Sem {\edge_a} = E_a\\
			\Sem{\edge_a\I} = E_a\I  \\ 
			\Sem{P/P'} = \Sem{P} \circ \Sem{P'}  \\ 
			\Sem{P^+} = \Sem{P}^+  \\ 
			\Sem{P\cup P'}  = \Sem{P}  \cup \Sem{P'} \\
			\Sem{\Goto(F)}  = \{ (v,v') \mid v' \in \Sem{F} \}
			\end{array}
			$
		\end{minipage}
	\end{minipage}
	\caption{\label{semantics} Semantics of \NRPQs on a $\Sigma$-labeled digraph
		$G=(V,(V_a)_{a\in \Sigma},(E_a)_{a\in\Sigma})$.}
\end{figure*}

The semantics of paths $P$ on labeled digraphs $G$ 
is the binary relation $\Sem{P}\subseteq V\times V$
defined in \Fig{semantics} in mutual recursivion with
the semantics of filters $\Sem{F}\subseteq V$.
Despite its binary semantics, we will use paths for 
defining sets of nodes by fixing a start set $S$ for the
navigation. So let $G$ be a labeled graph and $\START$ 
a subset of the  nodes of $G$.
For any $P\in\PA$, the set 
$\Sem{P}(\START)=   \{ v\mid \exists v'\in \START.\ (v',v) \in
  \Sem{P}\}$
contains all nodes that can
be reached when starting at some node of the start set $S$ 
and navigating over the path $P$. Similarly, the set 
$\Sem{F}(\START)= \Sem{F} \cap \START$
contains all  nodes from $S$
that satisfy the filter $F$.

\section{Top-Down Needed Subgraphs}
\label{sec:tdn}

We are interested in the top-down
evaluation of path queries, starting with a set of start nodes, and 
navigating along the path to other sets of nodes.
The \emph{top-down needed subgraph} of a path query
will be the subgraph visited by 
such a traversal.

For the formal definition, we consider labeled graphs as extensional
databases, i.e., as the sets of relational facts constructed from
a relational signature and a set of constants. More concretely,
we map any $\Sigma$-labeled graph
$G=(V,(V_a)_{a\in\Sigma},(E_a)_{a\in\Sigma})$ 
to the following set of database facts:
$$
\begin{array}{l} 
 \DLG(G) = \{ \nod(v)\mid v\in V\} \cup  \{ \nod[a](v)\mid v\in V_a,\ a\in\Sigma\} 
                            \\ \qquad \qquad \cup \ \{ \edg{a}(v,v')\mid (v,v')\in E_a,\ a\in\Sigma\}. 
\end{array}
$$

The facts are build from the monadic predicates $\nod$ and $\nod[a]$ and
the binary predicates $\edg{a}$ for all $a\in\Sigma$, and the graph
nodes $v\in V$ as constants. Conversely, consider
a set of facts $D$ with the following properties:
1. if $\nod[a](v)\in D$ then $\nod{(v)}\in D$ and 2. if $\edg{a}(v,v')\in
D$ then $\nod{(v)}\in D$ and $\nod{(v')}\in D$. For any such set $D$ there
exists a unique graph $G$ such that $\DLG(G)=D$. We
can therefore identify any graph $G$ with the
sets of facts $D=\DLG(G)$.

For any $\Sigma$-labeled digraph $G$ and set of start nodes $\START$
we define in \Figure{fig:tds} the set of \emph{facts of top-down needed subgraph}
 $\tdnGS(P)$ and $\tdnGS(F)$ for negation-free paths $P$ and
 filters $F$ in mutual recursion.
\begin{figure*}[t]
\begin{minipage}{\textwidth}
\begin{minipage}{.5\textwidth}
	$
	\begin{array}{l} 
          \tdnGS(\nod) =  \{\nod(v)\mid v \in \START\} \\
	\tdnGS(\nod[a]) =  \{\nod(v)\mid v\in\START\}\\
         \quad \cup \ \{\nod[a](v)\mid v\in V_a\cap\START\}            \\
 	\tdnGS(\filtertopath{F}) = \tdnGS(F)	  \\	
	\tdnGS(\edg{a}) =  \{\nod(v)\mid v\in\START\}\\ 
             \quad \cup \ \{\edg{a}(v,v'), \nod(v') \mid v\in
          \START,(v,v')\in E_a\}\\
    \tdnGS(\edg{a}\I) =   \{\nod(v)\mid v\in\START\}\\
           \quad \cup \ \{\edg{a}(v',v),\nod(v)\mid v'\in
          \START,(v,v')\in E_a\}\\ 
	\end{array}
$
\end{minipage}
\begin{minipage}{.49\textwidth}
$
        \begin{array}{l}	
	\tdnGS([P]) = \tdnGS(P)       \\
 \tdnGS(F\wedge F') = \tdnGS(F) \cup \tdnGstart{\Sem{F}(\START)}(F')\\
  \tdnGS(F\vee F') = \tdnGS(F) \cup \tdnGstart{\START}(F')\\
	\tdnGS(P/P') = \tdnGS(P) \cup \tdnGstart{\Sem{P}(\START)}(P')	\\	
	\tdnGS(P^+) = \tdnGstart{\Sem{P^+}(\START)}(P)                       \\	
	\tdnGS(P \cup P') = \tdnGS(P) \cup \tdnGS(P')	\\
 \tdnGS(\goto(F)) = \tdnG(F)  \qquad \textrm{(see \Figure{fig:tds-goto})}
	\end{array}
	$
\end{minipage}
\end{minipage}
\caption{\label{fig:tds} Facts of top-down needed subgraphs for
  negation-free
paths and filters.}
\end{figure*}
In the case of goto expressions, \Figure{fig:tds-goto} defines
$\tdnGS(\goto(F))=\tdnG(F)$ for restarting the computation with all nodes satisfying 
$F$.
\begin{figure*}
\begin{minipage}{\textwidth}
\begin{minipage}{.5\textwidth}
	$
	\begin{array}{l}
	\tdnG(\True) =  \{\node(v) \mid v\in V\}\\	
	\tdnG(\nodea) =  \{\nodea(v) \mid v\in V_a\} \\
 	\tdnG(\filtertopath{F}) = \tdnG(F) 	\\	
        \tdnG(\edge_a) =  \{\edgea(v,v')\mid (v,v')\in E_a\} \\
        \tdnG(\edge_a\I) =  \{\edgea(v',v)\mid (v,v')\in E_a\}
	\end{array}
$
\end{minipage}
\begin{minipage}{.49\textwidth}
$
        \begin{array}{l}	
\\
	\tdnG([P]) = \tdnG(P) \\			
   \tdnG(F\wedge F') = \tdnG(F) \cup  \tdnGstart{\Sem{F}}(F')\\
    \tdnG(F\vee F') = \tdnG(F) \cup \tdnG(F')\\
	\tdnG(P/P') = \tdnG(P) \cup \tdnGstart{\Sem{P}(V)}(P')
	\\	
	\tdnG(P^+) = \tdnGstart{\Sem{P^+}(V)}(P)
	\\	
	\tdnG(P \cup P') = \tdnG(P) \cup \tdnG(P')
	\\
	\tdnG(\goto(F)) = \tdnG(F) 
		\end{array}
	$
\end{minipage}
\end{minipage}
\caption{\label{fig:tds-goto} Top-down needed subgraphs without
  start sets as neeeded for goto expressions.}
\end{figure*}
%
The natural algorithm for computing the answer set of filter $\nodea$
at start set $\START$ will filter for all nodes $v\in\START$ such that
$v\in V_a$. Therefore all nodes in $\START$ need to be visited,
as well as the $a$-label of all nodes in $V_a\cap\START$. 
The extensional database of the top-down needed subgraph
$\tdnGS(a)$ therefore contains the facts in $\{\nod(v)\mid
v\in\START\}$ and $\{\nod[a](v)\mid v\in V_a\cap\START\}$.
The definition of $\tdnGS(F\wedge F')$ is
 sequential from the left to the right. When the filter query $F$ is failing
for a node $v$ then there is no need to check the filter query $F'$ so as to know
that the filter query $F\land F'$ is not verified by $v$. 
In contrast, the definition of $\tdnGS(F\lor F')$  is done a parallel manner, so that
both subfilters need to be evaluated from the start nodes.
The sequential alternative would lead to smaller  top-down needed subgraphs, which might seem advantageous:
$$
        \begin{array}{l}	
	  \tdnGSseq(F\vee F',\START) = \tdnGS(F) \cup \tdnGstart{\Sem{\neg  F}(\START)}(F').
	\end{array}
$$
However, obtaining an evaluator with this sequential behavior by compilation to 
datalog would require us to use stratified negation, that we prefer to 
avoid for the sake of presentation. For the same
reason, we restrict the definition of top-down needed
subgraphs to negation-free path queries.

\ignore{The filter query $F$ succeeds for $v$, we that the filter query $F\lor F'$ also succeeds for
$v$ without having to evaluate $F'$ at $v$. Of course, a parallel semantics
for these operators and visit a larger subgraph with the possibility of being
more time-efficient thanks to parallelism. We wish however to show that we can
visit an as small part of the graph as possible and so opt for sequential
logical operators.}

 The definition of $\tdnGS(P^+)$ is made of every
attempt to construct a path of $P$ starting from the nodes of $\START$ or the
nodes that can be reached from $\START$ with a path of $P^+$. 
In the case of goto expressions, we have defined
$\tdnGS(\goto(F))=\tdnG(F)$ for restarting the computation with all nodes satisfying 
$F$. We could set $\tdnG(F)$ to $\tdnGstart{V}(F)$,
but this would not be optimal since all nodes of $V$ 
would be top-down needed even for most simple filter $F=\nod[a]$.
A better definition where only the nodes of $V_a$ are top-down
needed is given in \Figure{fig:tds-goto}.

\begin{example}
	Consider the query $P_0 =\edge_a[\edge_b/\edge_c]$ on the graph $G_0$
        with signatue $\Sigma_0=\{a, b, c\}$ in \Fig{graph} with the
        start set $\START_0=\{0\}$. The set of
      top-down needed facts $\tdnGgen{G_0,\START_0}(P_0)$ is then
       $ \{ \edg{a}(0, 1),$ $\edg{a}(0, 4), $ $\edg{a}(0, 6),$
        $\edg{b}(1, 2),$ $\edg{b}(4, 2),$ $\edg{c}(2, 3)\}$.
    The top-down needed subgraph  
    which is annotated in red in \Fig{graph} is 
   thus $graph(\tdnGgen{G_0, \{0\}}(P_0))$ $=$ $(\{0, \ldots, 6\},$ $(V_\ell)_{\ell\in\Sigma_0\}},$ $ (E_\ell)_{\ell\in\Sigma_0})$ where $V_a=V_b=V_c=\emptyset$,
   $E_a=\{(0, 1),$ $(0, 4),$ $(0, 6)\}$, $E_b=\{(1, 2),$ $(4, 2)\}$, and $E_c=\{(2, 3)\}$.  
\end{example}

\section{Datalog Queries}\label{sec:datalog}

We recall preliminaries on the syntax and semantics of datalog programs
without negation and how to use them to define datalog queries on
extensional databases.

The syntax of datalog is parametrized by a finite set of 
predicates $p,q,r \in \Preds$ and a disjoint finite set of 
\emph{constants} $a, b, c\in\Consts$. The set of
predicates is partitionned into a subset  
of \emph{extensional predicates} $\ePreds$
and a 
disjoint subset of \emph{intensional predicates} $\iPreds$, 
so $\Preds=\ePreds\cup\iPreds$. 
Constants will serve as database elements and
extensional predicates for naming database relations.
An \emph{(extensional) database} is a subsets of 
ground literals of the form $p(a_1,\ldots,a_n)$ where $p\in\ePreds$ has 
arity $n\ge 0$ and $a_1,\ldots,a_n\in\Consts$.

We fix a set of variables $\Vars=\{x,y,z,\ldots\}$ distinct
from the constants and predicates. A \emph{term} 
$u,s,t \in \Terms=\Vars\uplus\Consts$ is either 
a variable or a constant. The set of \emph{(positive) literals} $\Lits$ 
is a subset of terms of the form
$q(u_1,\ldots, u_n)$ where $q\in\Preds$
has arity $n$ and $u_1,\ldots, u_n\in\Terms$.  
A vector of terms is denoted by $\ts\in\Terms^*$. 
%
The set of all literals
with extensional predicates is denoted by ${\eLits}$ and 
those with intensional predicates by ${\iLits}$.
A \emph{goal} is a vector of literals $\Ls\in\Lits^*$ that is to be
understood as a conjunction. The set of free variables $\fv(\ts),
\fv(\Ls)\subseteq \Vars$ are defined as usual. Similarly for the sets of
occuring constants $\consts(\ts),\consts(\Ls)\subseteq\Consts$.
A \emph{clause} is a pair of the form $q(\ts) \DL \Ls.$ where
$q(\ts)\in\iLits$ and $\Ls \in \Lits^*$. We call $q(\ts)$ 
the \emph{head} and
$\Ls$ the \emph{body} of the clause. The clause $q(\ts) \DL \Ls.$ is \emph{safe}
if $\fv(\ts)\subseteq\fv(\Ls)$. 
We only work with safe clauses throughout this paper.

A \emph{(safe) datalog program} is a finite subset $M$
of safe clauses. A\emph{ (safe) datalog query} has the form
$\DLQ\Ls{M}$, where  $\Ls\in\Lits^*$ is a datalog goal
and $M$ a safe datalog program $M$.
%
%
%
We now turn our attention to the semantics of datalog queries. 
Given a datalog query $\DLQ\Ls{M}$ and
an extensional database $D$, we need to define the set of 
substitutions that answer the query.
A \emph{substitution} is a finite partial function $\subs$ from $\Vars$ to
$\Terms$. We write $\lbrack\rbrack$ for the empty substitution.
Any substitution can be lifted to a total function on all variables
by defining $\sigma(x)=x$ for 
all $x\not\in\dom(\sigma)$. We lift substitutions 
further to total functions $\sigma:\Terms^*\to \Terms^*$ such that for 
all $n\ge 0$, $t_1,\ldots, t_n\in \Terms$ and
$a\in\Consts$:
$$
\begin{array}{lll}
\subs(t_1\ldots t_n) = \subs(t_1)\ldots \subs(t_n) &\text{and}&
\subs(a)=a
\end{array}
$$
Similarly, substitutions are lifted to functions
$\sigma:\Lits^*\to\Lits^*$ 
such that for all $\ts\in\Terms^*$ and $\ell_1$, $\ldots$, $\ell_n\in \Lits$:
$$
\begin{array}{lll}
      \subs(q(\ts))=q(\subs(\ts)), & \text{and}&
       \subs(\ell_1\ldots \ell_n)=\subs(\ell_1)\ldots \subs(\ell_n)  
\end{array}
$$

The \emph{renaming closure} of a program is the set of all clauses
that can be obtained from the clauses of the program by renaming 
variables bijectively:
$$
\begin{array}{l}
\ren(M) = \{ \subs(\ell) \DL \subs(\Ls) \mid \ell\DL\Ls. \text{ in }
  M,
  \subs \text{ is one-to-one substitution, } \ran(\subs)\subseteq \Vars \}
\end{array}
$$
We define joins and projections on 
substitutions as for the relational algebra: for any two substitutions $\subs$ and $\subs'$
and any finite subset of variables $V\subseteq \Vars$:
$$
\begin{array}{l@{\qquad}l}
        \subs \join \subs' = \left\{
   \begin{array}{ll}
      \subs \cup \subs'  &\text{if }\subs\cup \subs'\text{ is functional}
    \\
      \text{undefined} & \text{otherwise}
   \end{array}\right.
  &
        \Pi_{V}(\subs) = \subs_{|V}
\end{array}
$$


For any two literals $\ell,\ell'$
we define $\unif(\ell,\ell')$ as the most general unifier $\subs$ 
 such that $\subs(\ell)=\subs(\ell')$ if it exists, and leave it
undefined otherwise.

We define the semantics $\semM{\Ls}$ of a datalog query $\DLQ\Ls{M}$
on an extensional database $D$ as the least fixpoint 
that satisfies the equations in \Figure{fig:datalog-semantics}.
%
\begin{figure*}
	$
	\begin{array}{l}
		\semM{\epsilon}=\{\lbrack\rbrack\} \\
		\semM{\ell}=\left\{\begin{array}{ll}\{\Pi_{\fv(\ell)}(\subs\join \subs') \mid
		\subs=\unif(\ell,\ell'), \ \ell'\DL \Ls.
		\text{ in }\ren(M),\ \subs'\in\semM{\subs(\Ls)}\}&\text{if $\ell \in
			\iLits$}  \\
\{\Pi_{\fv(\ell)}(\subs) \mid \subs=\unif(\ell,\ell'), \ \ell'\in D\} &\text{if $\ell \in
			\eLits$}  
\end{array}\right.\\
		\semM{\ell_1\ldots \ell_n}=\{ \subs'\join \subs \mid \subs\in\semM{\ell_1}, \ \subs'\in
		\semM{\subs(\ell_2\ldots \ell_n)}\} \qquad \text{where }
          n\ge 2\\
	\end{array}
	$
	\caption{\label{fig:datalog-semantics} Least fixed-point semantics of
          a datalog query $\DLQ\Ls{M}$
		on a database $D$ for $\ell,\ell_1$, $\ldots$ $\ell_n$ $\in \Lits$.}
\end{figure*}
%
%
Notice that whenever we use the operation $\subs\join\subs'$
then we have $\dom(\subs) \cap \dom(\subs') = \emptyset$, so that
$\subs\join\subs'=\subs\cup\subs'$ is a well-defined
substitution. Each query answer $\sigma\in \semM{\Ls}$ has domain $\fv(\Ls)$ 
and always maps to constants since we work with safe datalog 
programs, so $\sigma:\fv(\Ls)\to\Consts$.
%
The semantics that we have given mimics the top-down 
datalog evaluation, which starts with the goal in the query and generates
subgoals by unfolding the clauses of the datalog program, while
instantiating the variables, until it reaches some ground facts 
from the extensional database. In general, this process 
may enter into infinite loops if not controlled by memoization.
The whole top-down evaluation can always be 
represented as a join tree as we illustrate by example in
\Figure{fig:td_eval_tree}. In the
case of infinite loops, the join tree is infinite. 

\ignore{This problem is avoided in datalog by using \colorbox{red}{memoization}: each
time a (variant up to renaming of variables of a) subquery is evaluated in the
process it is tabled allowing to reuse computation and detect loops. Making the
assumption that this \colorbox{red}{memoization} allows us to store and access each data in
constant time (for example, by using hash tables, or arrays if constants can be
aliased to integers), then the complexity of executing top-down evaluation is
linear in the number of solutions of  subqueries it finds
(see~\cite{tekle10:_precis_datal}). Here a solution of a subquery is understood
as the instantiation of a prefix of the body of a clause or of the query during
the evaluation. In top-down evaluation, negated goals are
then matched with failure to obtain the fact. In that case, this evaluation
mechanism computes the stratified semantics mentioned above.}

\begin{figure*}[t]
  \includegraphics[width=.8\textwidth]{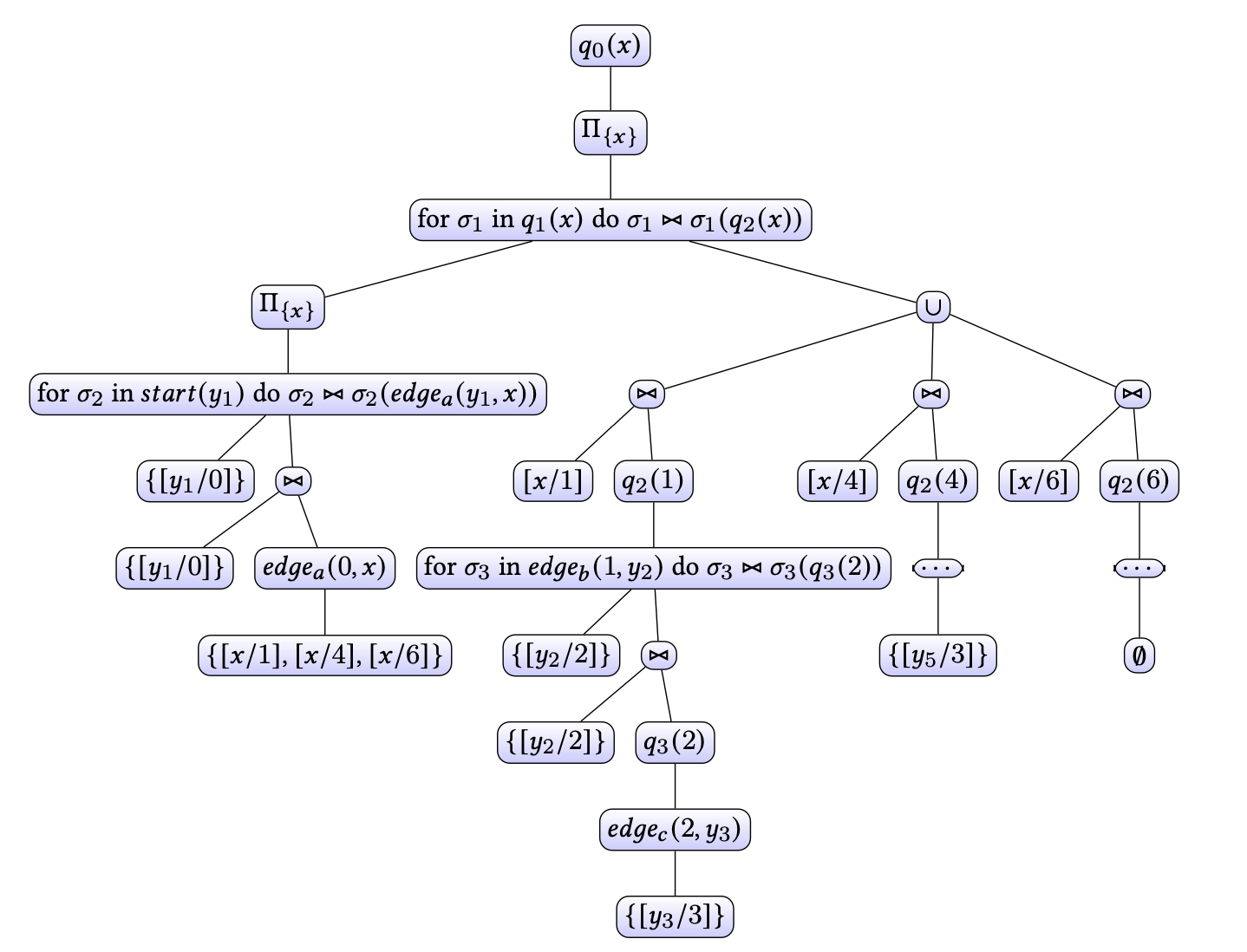}
	
	\caption{\label{substree}Top-down evaluation
		of	$\Trans{M_0,\DLG(G_0)\cup\{ start(0)\}}{q_0(x)}=\{[x/1],[x/4]\}$
                where $M_0$ is the datalog program from
                \Figure{datalog-for-P} 
                for
                $P_0 =\edge_a[\edge_b/\edge_c]$, and $G_0$ the
                graph from \Figure{graph}.}
	\label{fig:td_eval_tree}
\end{figure*}

\section{Complexity of Top-Down Evaluation of Datalog Queries}
\label{sec:complexity}

Known results on the complexity of top-down datalog evaluation
give us the formal tools
to prove for particular datalog queries, that the complexity 
of the top-down evaluation is in combined linear time
but with respect to the top-down visited sub-database, 
rather than with respect to the full database.

For any datalog query $\DLQ\Ls{M}$ and 
extensional database $D$ we next define 
the part of $D$ that is visited by the 
top-down evaluation of the datalog query. For this we assume 
that the set of extensional predicates of $D$ contains a monadic
predicate $\node\in\ePreds$ such that $\node^D=\Consts$.
We define the \emph{top-down visited sub-database} $\VisM(\Ls)$ as
the extensional database over $\ePreds$ -- following  the semantics 
of datalog queries -- as the least fixed point of equations in \Figure{fig:complexity-visited}.

\begin{figure*}
$
\begin{array}{l}
\VisM(\epsilon) = \emptyset\\
  \VisM(\ell)=
\left\{\begin{array}{@{\hspace{-.05cm}}l@{\hspace{0.1cm}}l}  
\{ \node(a)\mid a \in \consts(\ell)\} \cup 
\bigcup \left\{  \ell' \mid
  \unif(\ell,\ell') \text{ defined}, \ell' \text{ in }D \right\}
  & \text{if } \ell\in\eLits\\
\{\node(a)\mid a \in \consts(\ell)\}
                  \cup \bigcup \{ \VisM(\subs(\Ls))
  \mid   \subs=\unif(\ell,\ell'),\,\ell'\DL \Ls. \text{ in }\ren(M)
\} &\text{if } \ell\in\iLits
\end{array}\right.\\
\VisM(\ell_1\dots\ell_n) = \VisM(\ell_1) \cup \bigcup\{
  \VisM(\sigma(\ell_2\dots\ell_n))\mid \sigma\in\semM{\ell_1}\}\\

\end{array}
$
\caption{\label{fig:complexity-visited} The top-down visited sub-database $\VisM(\Ls)$ where $\ell, \ell_1$,\dots,$\ell_n\in\Lits$ and $n\ge 2$.}
\end{figure*}


\begin{definition}\label{def:SCL}
  We call a datalog goal $\Ls$ \emph{simply combined
    linear} (SCL) if any proper prefix of $\Ls$ is SCL and 
  $\fv(\Ls)$ is either guarded by a single extensional literal of $\Ls$ or
  contains no more than one variable. We call 
a datalog query $\DLQ{\Ls}{M}$ SCL if the datalog goal $\Ls$ is SCL
and for each of the clauses $\ell\DL\Ls.$ in the datalog program $M$,
 the datalog goal $\ell\Ls$ is SCL.
\end{definition}
For example, let $p$, $q\in\iPreds$ be monadic and
$r\in\ePreds$ be binary. The goal $p(x),r(x,y),$ $q(y)$
is then SCL, since its prefix $p(x)$ contains no more than
one variable, and both of its variables $x$ and $y$ are guarded
by the extensional literal $r(x,y)$. The
goal $p(x),r(x,x),q(y)$ on the contrary is not SCL,  as it 
contains two variables of which $y$ is not guarded by
any extensional literal. The goal $p(x),q(x)$
is SCL since it contains no more than a single free
variable. 

Given an extensional
database $D$, any SCL goal $\Ls$ has a number of 
ground instances that is linear in the size of $D$. Even better
the number of ground instances inspected by top-down
evaluation of the datalog query $\DLQ\Ls{M}$ is linear 
in the size of the top-down visited database $\VisM(\Ls)$.
In the case where $\fv(\Ls)$ contains at most one variable, 
this variable must be instantiated by some node of
the top-down visited sub-database. Otherwise,
the set of free variables $\fv(\Ls)$ is guarded by a single 
extensional literal of $\Ls$, say $p(\ts)\in\eLits$. In this case,
any ground instance of $\Ls$ visited by the top-down
evaluation of $M$ is determined by $\unif(p(\ts),p(\vs))$
for some fact $p(\vs)\in\VisM(\Ls)$. 

\begin{theorem}
  \label{theorem-top-down-visited}
  The answer set $\semM{\Ls}$ of a safe SCL query
  $\DLQ\Ls{M}$ on an extensional database $D$ can be computed in time
  $\mathcal{O}(|M||\VisM(\Ls)|)$.
\end{theorem}

For proof we can show for any safe SCL datalog
  queries, that its top-down evaluation with memoization can be
  done in combined linear time with respect to the size of the
  top-down visited sub-database. For this, we can rely on the
  top-down evaluator in Figure 1 of \cite{tekle10:_precis_datal}. The needed arguments on
safe SCL datalog programs were given above. 
We also note that the magic set transformation on datalog programs
without negation can be used to reduce top-down evaluation
with memoization to semi-naive bottom-up evaluation.
As stated by Theorem 3 of Ullman \cite{Ullman89},
the bottom-up evaluator obtained is at least as
time efficient as the top-down evaluator. The magic
set transformation, however, may need exponential space. Tekle and Liu
\cite{tekle10:_precis_datal} show that this problem can be solved by
perfoming it on demand. They also proposed an on demand
magic set transformation for stratified
datalog programs \cite{Tekle_2019}.

\section{Compiler to SCL Datalog Queries}
\label{sec:compiler}

We now contribute the compiler from negation-free path queries $P$
and start set $\START$ to SCL datalog queries $\DLQ\Ls{M}$, such that
for any graph $G$ with nodes subsuming $\START$, the 
extensional database of the top-down needed subgraph 
$\tdnGS(P)$ is equal to the top-down visited sub-database 
$\Vis_{M,\DLG(G)}(\Ls)$. The top-down evaluation of the
datalog query $\DLQ\Ls{M}$ on the graph's database $\DLG(G)$
thus yields the expected upper complexity bound for the 
evaluation of path queries by Theorem \ref{theorem-top-down-visited}.

For any set of start nodes $\START$ and monadic
predicate $i\in\iPreds$, we define a datalog program 
$\DLS{i} = \{i(v)\DL. \mid v \in \START\}$.
The compilation scheme for path queries 
follows the structure of paths and filters
by mutual recursion. It is given by the
datalog programs $\ACC{P}{i,f}$ in \Figure{compiler_monadic_acc},
$\FILT{F}{c}$ in \Figure{compiler_monadic_filters} and
$\EX{P}{c,r}$ in \Figure{compiler_monadic_ex}. 
Path queries outside filters need to compute
all accessible nodes by $\ACC{P}{i,f}$, while path queries 
within filters need to check the existence of 
accessible nodes by $\EX{P}{c,r}$.
The compiler introduces fresh monadic predicates 
for all subexpressions: \emph{initial predicates}
$i,i',i''\in\iPreds$, \emph{final predicates} 
$f, f',f''\in\iPreds$ \emph{final}, \emph{checks}.  
$c,c',c''\in\iPreds$, and continuations  $r,r',r''\in\iPreds$.

 Given a graph $G$ and with a start set 
$\START\subseteq V$ of graph nodes, the answer set of the
datalog query $\DLQ{f(x)}{\ACC{P}{i,f} \cup \DLS{i}}$ on the
extensional database $\DLG(G)$ is $\{[x/v]\mid v\in\Sem{P}(\START)\}$, 
assigning the free  variable $x$ to some
node $v$ reachable from $\START$ over $P$ in $G$. The
initial predicate $i$ captures the set of start nodes, and
the \emph{final}  predicate $f$
the answer set of the path query $P$ started from there.
The fresh monadic predicates make the datalog programs 
for the subexpressions able to communicate. For instance, 
we have $\ACC{P'/P''}{i,f} = \ACC{P'}{i,f'} \cup \ACC{P''}{f',f}$. Here the
final predicate $f'\in\iPreds$ represents the answer 
set of path $P'$ started at node set $i$, but also
the start set for the path $P''$. This is since the start nodes of 
$P''$ in the query $P'/P''$ are the nodes that are reached with the
query $P'$.
For the recursive path queries $P^+$ we have $\ACC{P^+}{i,f}
= \ACC{P}{i,f}\cup\{i(x)\DL f(x).\}$. Here the rule $i(x)\DL f(x).$ represents
the fact that once a node is reached by the query $P^+$ it becomes a
possible start node for the same query. 

We next consider the datalog programs $\FILT{F}{c}$ defined
in \Figure{compiler_monadic_filters}. For any graph $G$ 
the answer set of the datalog query $\DLQ{ c(x)}{\FILT{F}{c} } $ 
on the extensional database $\DLG(G)$ is 
$\{[x/v]\mid v \in \Sem{F}\}$, so that the 
free variables $x$ may be bound to any
node seleced by the filter. Hence, for any
start set $\START$, the answer set of 
$\DLQ{ i(x),c(x)}{\FILT{F}{c} \cup \DLS{i}} $ 
is $\{[x/v]\mid v \in \Sem{F}(\START)\}$.
The filter for all nodes is compiled to
$\FILT{\True}{c} = \{c(x)\DL \node(x)\}$. Thereby, the check $c$
is called for all nodes of the graph. Note that $\node$ is an
extensional predicate, so this clause is safe. A conjunction
of filters $\FILT{F'\wedge F''}{c}$ is compiled by adding the 
clause $c(x) \DL c'(x), c''(x)$ to the datalog programs 
$\FILT{F'}{c'}$ and  $\FILT{F''}{c''}$. The added clause 
checks sequentially, whether a node $x$ is filtered by $F'$ and 
if so whether it is also filtered by $F''$. A disjunction
of filters $\FILT{F'\vee F''}{c}$ is compiled by adding the 
two clause $c(x) \DL c'(x).$ and  $c(x)\DL c''(x).$ to the datalog programs 
$\FILT{F'}{c'}$ and  $\FILT{F''}{c''}$. The two added clauses 
check in parallel whether a node $x$ is filtered by $F'$ 
or whether $x$ is filtered by $F''$.

\begin{toappendix}
Our compiler could
be extended to negated filters, but this would require
to compile to stratified datalog. Startified negation
would also allow to compile disjunctions sequentially, 
so that only a smaller subgraph would get visited.

The accessiblity of nodes over goto-paths could be expressed by
 $$ \ACCTILDE{goto(F')}{i,f} = \FILT{F'}{f'} \cup \{f(x)\DL i(y), f'(x).\}$$
where $\FILT{F'}{f'}$ is the datalog program that represents the
computation of the filter query $F'$ and where $f'$ is the predicate which
captures the answer set. The clause $f(x)\DL i(y), f'(x).$ means that
a node $x$ satisfying the filter $F'$ is in the answer set 
if there is some node $y$ in the start set $\START$. 
So if $\START\not=\emptyset$ then the datalog query 
$\DLQ{f'(x)}{\FILT{F'}{c}\cup\DLS{i}}$ can be reduced
to the datalog query $\DLQ{c(x)}{\FILT{F'}{c}}$, which
on the extensional database $\DLG(G)$ 
has the answer set $\{[c/v]\mid
v\in\Sem{F'}\}$. We note that the top-down evaluation of the 
latter datalog query avoids visiting the nodes of the graph that
are not top-down needed for the path $goto(F')$.  If for instance $F'=\nodea$
then only the $a$-labeled nodes of the graph are visited.
The problem with the definition discussed so far,
however, is that the clause $f(x)\DL i(y), f'(x)$ is not SCL 
(see Definition \ref{def:SCL}). Even worse, 
it would lead to a quadratic evaluation time. Therefore,  
we replace it 
by the equivalent datalog program with two clauses $\{f(x)\DL j(), f'(x).\}$ $\cup$
$\{j()\DL i(y).\}$ which is SCL, leading to the definition of $\ACC{goto(F')}{i,f} $
in \Figure{compiler_monadic_acc}. Here $j\in\iPreds$ is a 
fresh nullary predicate that is true for $\DLS{i}$ if and only if $\START\not=\emptyset$.
\end{toappendix}

In \Figure{compiler_monadic_ex} we define the
datalog programs $\EX{P}{c,r}$ for evaluating paths
$P$ existentially as needed when paths are used in filters,
that is $\FILT{\lbrack P \rbrack}{c} =  \EX{P}{c,r} \cup \{ r(x) \DL node(x).\}$. 
The check predicate $c$ denotes the set of source nodes, 
from which some target node can be reached over $P$, while $r$ is the
continuation to which the target node must belong.
Given a graph $G$ and a start set $\START$, the answer set of 
the datalog query $\DLQ{c(x)}{\EX{P}{c,r}  \cup \{r(x)\DL node(x).\}}$ on the extensional database 
$\DLG(G)$ is $\{[c/v] \mid (v,v')\in\Sem{P}\}$. The continuation
predicate $r$ is required to allow us to compile path concatenations in
filters, i.e., in $    		\EX{P'/P''}{c,r} = \EX{P'}{c,f} \cup \EX{P''}{f,r}$. 
Note that the interplay of the predicate $c$ and $r$ is similar to the
one between $i$ and $f$ in $\ACC{P}{i,f}$.

\begin{figure*}
	
 \noindent\begin{minipage}{\textwidth}
	    \begin{minipage}{.5\textwidth}
		$
		\begin{array}{l} 
		\ACC{\edge_a}{i,f} =  \{f(x) \DL  i(y), \edge_a(y,x).\} \\ 
		\ACC{\edge_a\I}{i,f} =  \{f(x) \DL i(y), \edge_a(x,y).\} \\ 
		\ACC{P'/P''}{i,f} = \ACC{P'}{i,f'} \cup \ACC{P''}{f',f} \\
		\ACC{P^+}{i,f} = \ACC{P}{i,f}\cup\{i(x)\DL f(x).\} \\
		\end{array}
		$
	    \end{minipage}
	    \begin{minipage}{.49\textwidth}
		$
		\begin{array}{l}	
		\ACC{P' \cup P''} {i,f}=\ACC{P'}{i,f}\cup\ACC{P''}{i,f} \\
		\ACC{goto(F')}{i,f} = \FILT{F'}{f'} \ \cup \\ \qquad \{f(x)\DL j(), f'(x). \quad j()\DL i(x).\} \\
		\ACC{\filtertopath{F'}}{i,f} =  \FILT{F'}{f'} \cup \{ f(x) \DL i(x), f'(x).\} \\
		\end{array}
		$
	    \end{minipage}
    \end{minipage}
    \caption{\label{compiler_monadic_acc}The datalog program
      $\ACC{P}{i,f}$ for path $P$ and monadic predicates
      $i,f\in\iPreds$. }

\medskip
\noindent\begin{minipage}{\textwidth}
    	\begin{minipage}{.5\textwidth}
    		$
    		\begin{array}{l} 
    		\FILT{a}{c}  = \{ c(x)\DL \nodea(x). \}\\
    		\FILT{\True}{c} =  \{c(x) \DL node(x).\} \\ 
    		\FILT{F'\vee F''}{c} = \FILT{F'}{c'} \cup \FILT{F''}{c''} \ \cup \\ \qquad \{c(x)\DL
    		c'(x).\quad c(x)\DL 
 c''(x).\} \\
    		\end{array}
    		$
    	\end{minipage}
    	\begin{minipage}{.49\textwidth}
    		$
    		\begin{array}{l}	
    		\FILT{F'\wedge F''}{c} = \FILT{F'}{c'} \cup \FILT{F''}{c''} \ \cup \\ \qquad
    		\{ c(x) \DL c'(x), c''(x).\} \\
    		\FILT{\lbrack P \rbrack}{c} =  \EX{P}{c,r} \cup \{ r(x) \DL node(x).\} 
    		\end{array}
    		$
    	\end{minipage}
    \end{minipage}
    \caption{\label{compiler_monadic_filters}The datalog program
      $\FILT{F}{c}$ for filter $F$ and monadic predicate
    	$c\in\iPreds$. }
\end{figure*}

\begin{figure*}[t]
\noindent 
    \begin{minipage}{\textwidth}
    	\begin{minipage}{.5\textwidth}
    		$
    		\begin{array}{l} 
    		\EX{\edge_a}{c,r} =  \{c(x) \DL \edge_a(x,y), r(y).\} \\ 
    		\EX{\edge_a\I}{c,r} =  \{c(x) \DL \edge_a(y,x), r(y).\} \\ 
    		\EX{P'/P''}{c,r} = \EX{P'}{c,f} \cup \EX{P''}{f,r} \\
    		\EX{P^+}{c,r} = \EX{P}{c,r} \cup \{r(x) \DL c(x).\}\\
    		\end{array}
    		$
    	\end{minipage}
    	\begin{minipage}{.49\textwidth}
    		$
    		\begin{array}{l}	
    		\EX{P' \cup P''} {c,r}=\EX{P'}{c,r}\cup\EX{P''}{c,r}\\
    		\EX{goto(F')}{c,r} = \FILT{F'}{c'} \ \cup \\ \qquad \{c(x)\DL j(). \quad j()\DL c'(y), r(y).\}  \\
    		\EX{\filtertopath{F'}}{c,r} =  \FILT{F'}{c'} \cup \{ c(x) \DL c'(x), r(x).\}\\
    		\end{array}
    		$
    	\end{minipage}
    \end{minipage}
    \caption{\label{compiler_monadic_ex}The datalog program $\EX{P}{c,r}$ for path 
    	$P$ with monadic predicates $c, r\in\iPreds$. }

\end{figure*}

\begin{lemmarep}
  \label{lem:well-formedness}
  For any path $P$, filter $F$, graph $G$, start set $\START$, 
  and monadic predicates $i,f,c,r\in\iPreds$, 
  the programs $\DLS{i}$, $\ACC{P}{i,f}$,
  $\FILT{F}{c}$,  $\EX{P}{c,r}$ 
  are safe and SLC.
\end{lemmarep}

\begin{toappendix}
\begin{proof}
Elementary by inspection of all cases of the definitions of these
datalog programs.
\end{proof}
\end{toappendix}

The function
$\reached{M,r}(\Ls\ )$ defined in \Figure{compiler_reached}
returns the set of all nodes $v$, such
that $r(v)$ is queried in the proccess of the top-down evaluation of
the datalog query $\DLQ{\Ls}{M}$.
\begin{figure*}
$
\begin{array}{l}
\reached{M,r}(\epsilon)
= \emptyset \\
\reached{M,r}(r(v),\Ls_1)
=
\{v\}\cup  \reached{M,r}(\subs(\Ls_2,\Ls_1))
\mid  \subs=\unif(r(v),\ell'),  \ell'\DL \Ls_2. \text{ in }\ren(M)\}
\\
\reached{M,r}(\ell,\Ls_1)=
\reached{M,r}(\subs(\Ls_2,\Ls_1)) \mid 
\subs=\unif(\ell,\ell'),\
\ell'\DL \Ls_2. \text{ in }\ren(M)\} \quad\text{if } \ell \neq r(v)
\end{array}
$
\caption{\label{compiler_reached}The $\reached{M,r}$ function for the datalog query $\DLQ{\Ls}{M}$. }
\end{figure*}
Now, we provide the Propositions \ref{main_filt_and_ex} and \ref{main_acc} 
for dividing the correctness proof into two parts. First --- about subpaths and subfilters of
some filter. Concommittantly with
Theorem~\ref{theorem-top-down-visited} they will
imply the main efficiency Theorem \ref{theo:correct}.

\begin{propositionrep}\label{main_filt_and_ex}
	For any filter query $F \in \FI$, path query $P \in \PA$,
        label $a \in \Sigma$, labeled graph $G$, subset
        $\START\subseteq V$ of nodes
	of $G$, distinct monadic predicates $i,c,r\in\iPreds$ and
        $x\in\Vars$.
        
\noindent
     	\begin{minipage}{\textwidth}
	\begin{minipage}{.38\textwidth}
     \begin{enumerate}
		\item if $M=\FILT{F}{c}\cup\DLS{i} $ and $\Ls= i(x),c(x)$
		then: 
		\begin{itemize}
			\item $\Trans{M,\DLG(G)}{\Ls} = \{[x/v] \mid v \in  \Sem{F}(\START)\}$
			\item \mbox{$\Vis_{M,\DLG(G)}( \Ls) =
                            \tdnGS(F)$\hspace{-2em}}
	    \end{itemize}
          \end{enumerate}
        \end{minipage}
        \begin{minipage}{.62\textwidth}  
     \begin{enumerate}
		\item[2.] if $M=\EX{P}{c,r}\cup\DLS{i} \cup \{r(x)\DL node(x).\}$ and $\Ls= i(x),c(x)$ then:
		\begin{itemize}
			\item $\Trans{M,\DLG(G)}{\Ls} = \{[x/v]\mid v \in\START,\ 
			\Sem{P}(\{v\})\neq\emptyset\}$
			\item $\Vis_{M,\DLG(G)}(\Ls) = \tdnGS(P)$
			\item $reached_{M}(\Ls) = \Sem{P}(\START)$
		\end{itemize}
              \end{enumerate}
            \end{minipage}
            \end{minipage}
\end{propositionrep}

\begin{toappendix}
  \begin{proof}
	By simultaneous induction on the structures of 
      $F\in \FI$ and $P\in \PA$.
	We discuss a few cases for the possible forms of filters
      and paths only.
For warming up, we consider the proof of property (1) for filters
$F=F'\wedge F''$. The definition of the compiler yields:
		$$
		\FILT{F'\wedge F''}{c} = \FILT{F'}{c'} \cup \FILT{F''}{c''} \cup \{c(x) \DL c'(x), c''(x).\}
		$$
		 
		Let $M=\FILT{F'\wedge F''}{c}\cup\DLS{i}$, $M' =
                \FILT{F'}{c'}\cup\DLS{i}$, and $M'' =
                \FILT{F''}{c''}\cup\DLV{i}{\Sem{F'}(\START)}$. The
                top-down evaluator for $\DLQ{ i(x),c(x)}{M}$ will visit $\tdnGS(F')$ in order to
                get all facts corresponding to the filter query
                $F'$. After that, the top-down evaluator will start
                from all nodes in $\START$ that satisfy filter $F'$ to
                find those that also satisfy the filter $F''$.
		First, we show that the top-down evaluation
                $\Trans{M}{i(x),c(x)}$ yields $\{[x/v]\mid v\in \Sem{F'\wedge F''}(\START)\}$. By induction hypothesis, $$\Trans{M'}{i(x),c'(x)} = \{[x/v] \mid v \in \Sem{F'}(\START)\}$$ and $$\Trans{M''}{i(x),c''(x)} = \{[x/v] \mid v \in \Sem{F''}(\Sem{F'}(\START))\}.$$ Therefore, 
                $$
                \begin{array}{rcl}
                \Trans{M}{i(x),c(x)} &=& 
                \{[x/v] \mid v\in \Sem{F''} \cap (\Sem{F'}  \cap \START)\}
                \\
                &=&
                \{[x/v] \mid v\in \Sem{F' \wedge F''} \cap \START\}
                \\
                &=&
                \{[x/v] \mid v\in \Sem{F' \wedge F''}(\START)\}.
                \end{array}
                $$
		
		Second, we show that the top-down visited sub-database is equal to $\tdnGS(F'\wedge F'')$. By induction hypothesis, $$\Vis_{M',\DLG(G)}(i(x),c'(x)) = \tdnGS(F')$$ and $$\Vis_{M'',\DLG(G)}(i(x),c''(x)) = \tdnGstart{\Sem{F'}(\START)}(F'').$$ Therefore,
		$$
		\begin{array}{rcl}
		\Vis_{M,\DLG(G)}(i(x),c(x)) &=& 
		\tdnGS(F') \cup \tdnGstart{\Sem{F'}(\START)}(F'')
		\\
		&=&
		\tdnGS(F' \wedge F'').
		\end{array}
		$$
		
We now turn to the most complicated case, which is the proof of
property 2 for  paths $P = P'/P''$. The definition of the
compiler yields:
		$$
		\EX{P'/P''}{c,r} = \EX{P'}{c,f} \cup \EX{P''}{f,r}
		$$
		
		Let $M=\EX{P'/P''}{c,r}\cup\DLS{i} \cup \{r(x)\DL
                node(x).\}$, $M' =
                \EX{P'}{c,f}\cup\DLS{i} \cup \{f(x)\DL node(x).\}$,
    and $M'' = \EX{P''}{f,r}\cup\DLV{i}{R} \cup \{r(x)\DL node(x).\}$.
 We can divide the top-down evaluation of
                the datalog query $\DLQ{i(x),c(x)}{M}$ into two
                steps. First --- the top-down evaluation of the
                datalog query $\DLQ{i(x),c(x)}{M'}$ 
                reaches the set $R$ of facts of the form $f(v)$ where $R  = reached_{M',
                  f}(i(x),c(x))$. The second
                step is equivalent to the evaluation of the datalog
                query $\DLQ{i(x),f(x)}{M''}$ where the set $R$ is used as the set of starting nodes. By induction hypothesis, $R = \Sem{P'}(\START)$. Therefore, the result of top-down evaluation 
		$\Trans{M,\DLG(G)}{i(x), c(x)}$ is equal to the join of the result of evaluation $\Trans{M',\DLG(G)}{i(x), c(x)}$ and result of evaluation $\Trans{M'',\DLG(G)}{i(x), f(x)}$. By induction hypothesis,
		$$\Trans{M',\DLG(G)}{i(x), c(x)} = \{[x/v]\mid v \in\START,\ 
		\Sem{P'}(\{v\})\neq\emptyset\}$$ and $$\Trans{M'',\DLG(G)}{i(x), f(x)} = \{[x/v]\mid v \in R,\ 
			\Sem{P''}(\{v\})\neq\emptyset\}.$$ Let $\Ls = i(x), c(x)$. Therefore, 
		$$
		\begin{array}{rcl}
		\Trans{M,\DLG(G)}{\Ls} &=& 
		\{ \subs \join \proj_{\emptyset}(\subs') \mid \\ & & \ \subs \in \{[x/v] \mid v \in\START,\  
		\Sem{P'}(\{v\})\neq\emptyset \},\\ & & \ \subs' \in \{[x/v,y/v'] \mid v \in\START, \\ & & \  (v,v') \in \Sem{P'}, \Sem{P''}(\{v'\})\neq\emptyset\} \}
		\\
		&=&
		\{ [x/v] \mid v \in \START, \exists v',v''. (v,v') \in \Sem{P'}, \\ & & \  (v',v'') \in \Sem{P''} \}
		\\
		&=&
		\{ [x/v] \mid v \in \START, \ \Sem{P'/P''}(\{v\})\neq\emptyset \}.
		\end{array}
		$$
		
	    Next, we show that the top-down visited sub-database is
            equal to $\tdnGS(P'/P'')$. By induction hypothesis, we
            obtain $$\Vis_{M',\DLG(G)}(i(x), c(x)) = \tdnGS(P')$$ and
		$$\Vis_{M'',\DLG(G)}(i(x), f(x)) = \tdnGstart{\Sem{P'}(\START)}(P'').$$ Therefore,
		
		$$
		\begin{array}{rcl}
		\Vis_{M,\DLG(G)}(\Ls) &=& 
		\tdnGS(P') \cup \tdnGstart{\Sem{P'}(\START)}(P'')
		\\
		&=&
		\tdnGS(P'/P'').
		\end{array}
		$$
		
		Finally, we show that the set of all nodes $v$, such that the $r(v)$ is queried during the top-down evaluation of the datalog query $\DLQ{i(x),c(x)}{M}$, is equal to $\Sem{P'/P''}(\START)$. The $r(v)$ can be queried only in the second step of the top-down evaluation which is equivalent to the evaluation of the datalog query $\DLQ{i(x),f(x)}{M''}$. By induction hypothesis, $$reached_{M'', r}(i(x), f(x)) = \Sem{P''}(\Sem{P'}(\START)).$$
		Thus,
		$$
		\begin{array}{rcl}
		reached_{M, r}(i(x), c(x)) &=& 
		reached_{M'', r}(i(x), f(x))
		\\
		&=&
		\Sem{P''}(\Sem{P'}(\START))
		\\
		&=&
		\Sem{P'/P''}(\START).
		\end{array}
		$$
\end{proof}
\end{toappendix}

\begin{propositionrep}\label{main_acc}
	For any path query $P \in \PA$, labeled graph $G$, subset
        $\START$ of nodes of $G$, distinct intensional predicates
        $i,f\in\iPreds$ and $x\in\Vars$, 
        if $M=\ACC{P}{i,f}\cup\DLS{i}$ then:

        \noindent
	\begin{minipage}{\textwidth}
	\begin{minipage}{.6\textwidth}
\begin{itemize}
	\item  $\Trans{M,\DLG(G)}{f(x)} = \{[x/v] \mid v \in
          \Sem{P}(\START)\}$
        \end{itemize}
	\end{minipage}
	\begin{minipage}{3.95\textwidth}
        \begin{itemize}
	\item $\Vis_{M,\DLG(G)}(f(x)) = \tdnGS(P)$
        \end{itemize}
      \end{minipage}
      \end{minipage}
\end{propositionrep}

\begin{toappendix}
  \begin{proof}
	Let $G$ be a graph with a node set $V$.
	The proof is induction on the structure of paths $P\in \PA$.
	We only consider the case $P = P'/P''$.
		The definition of the compiler yields:
		$$
		\ACC{P'/P''}{i,f} = \ACC{P'}{i,f'} \cup \ACC{P''}{f',f}
		$$
		Let $M=\ACC{P'/P''}{i,f}\cup\DLS{i}$, $M' =
                \ACC{P'}{i,f'}\cup\DLS{i}$ and $M'' = \ACC{P''}{f',f} \cup \DLV{f'}{R}.\}$.
              We can divide the top-down evaluation of the datalog query $\DLQ{f(x)}{M}$ into two steps. 
	First --- the top-down evaluation of the datalog query $\DLQ{f'}{M'}$ provides the set of nodes $R$ equal to the set of all nodes $v$, such that $f'(v)$ is inferred after the first step of evaluation. The second step is equivalent to the evaluation of the datalog query $\DLQ{f(x)}{M''}$  where the set $R$ is used as the set of starting nodes. By induction hypothesis, $R = \Sem{P'}(\START)$. Therefore,	
	$$
	\begin{array}{rcl}
	\Trans{M,\DLG(G)}{f(x)} &=& 
	\Trans{M'',\DLG(G)}{f(x)} = \{ [x/v] \mid \\ & & \ v \in \Sem{P''}(\Sem{P'}(\START)) \}
	\\
	&=&
	\{ [x/v] \mid v \in \Sem{P'/P''}(\START) \}.
	\end{array}
	$$

		Next, we show that the top-down visited sub-database
                is equal to $\tdnGS(P'/P'')$. By induction hypothesis,
                we obtain $$\Vis_{M',\DLG(G)}(f'(x)) = \tdnGS(P')$$ and
		$$\Vis_{M'',\DLG(G)}(f(x)) = \tdnGstart{\Sem{P'}(\START)}(P'').$$
		Therefore,
		$$
		\begin{array}{rcl}
		\Vis_{M,\DLG(G)}(f'(x)) &=& 
		\tdnGS(P') \cup \tdnGstart{\Sem{P'}(\START)}(P'')
		\\
		&=&
		\tdnGS(P'/P'').
		\end{array}
		$$	
\end{proof}
\end{toappendix}

\begin{theoremrep}\label{theo:correct}
For any graph $G$ with subset of nodes $\START$ and 
any path query $P \in \PA$ the answer set $\Sem{P}(\START)$ 
can be computed in time $\mathcal{O}(|P|| \tdnGS(P)|)$.
\end{theoremrep}

\begin{appendixproof} 
This follows from Proposition \ref{main_acc}, Proposition \ref{main_filt_and_ex} and
Theorem~\ref{theorem-top-down-visited}. 
\end{appendixproof}



\section{Jumping in Graphs}
\label{sec:jumping}

Preprocessing is mandatory for sharing efforts when 
evaluating multiple queries on the same large graph. 
Most typically, one can pre-compute indexes that give 
efficient access to some particular relations of the graph. 
Here we consider indexes, which are binary relations 
defined by \NRPQs themselves.

For instance, we might want to jump from a node of the graph to 
the next $a$-labeled node in some fixed total order. 
In this case, one would like to have a jumping algorithm
that visits only the top-down needed subgraph, but 
taken with respect to the graph, that is enriched 
with extra edges labeled by the names of the
indexes.

\newcommand\Pex{acc_a}
Let us next consider a little more complex example. For this
we suppose that we have an index for the \NRPQ
$\Pex=\edge^*/\filtertopath{a}$. We can then extend the signature $\Sigma$ with
a new label $\Pex$, the graph $G$ with $\Pex$-labeled 
edges for all pairs in $\Sem{\Pex}$, and rewrite the target
path query by substituting all its subqueries $\Pex$ by 
$\edge_{\Pex}$. This has the advantage that fewer nodes 
are top-down needed after the rewriting on the enriched graph. 
For instance, a top-down evaluator for the path query $\Pex$ 
without jumping needed to inspect \emph{all} nodes of the
graph accessible from $S$, 
since all of them needed to be tested for whether they
satisfied the filter query $a$. After the rewriting to $\edge_{\Pex}$, 
a top-down algorithm can jump directly from the start nodes 
in $S$ to the accessible $a$-labeled nodes by using the index, so 
only accessible $a$-labeled nodes will be visited.

The general jumping algorithm starts with a set of 
indexes for \NRPQs say for $P_1,\ldots,P_n$. For
answering a query $P$ on a graph $G$ with these indexes the
jumping algorithm enriches the signature $\Sigma$ by new labels 
$P_1,\ldots,P_n$, the original graph $G$ with new labeled edges 
$E_{P_j}=\Sem{P_j}$ where $1\le j\le n$, and then substitutes 
in the target 
query $P$ all occurrences of the subqueries $P_j$ by $\edge_{P_j}$. The order of the substitution
can be chosen arbitrarily, depending on the
intended jumping strategy. In this way, the top-down needed subgraph of
the enriched graph for the rewritten query is intuitively 
exactly the subgraph of the original graph that a 
top-down evaluation algorithm with jumping 
needs to visit.

This  jumping algorithm can be used to reformulate in 
simple terms a variant of  the efficient  automata-based algorithm from
\cite{ManethNguyen10} that evaluates navigational path queries.
More precisely, their algorithm covers navigational forward \XPath
queries on \XML documents. It is based on alternating tree automata
with selection states, which can be seen a binary datalog programs,
while ours are monadic. \XML documents are seen as 
labeled graphs, with two edge labels \emph{firstchild}
and  \emph{nextsibling}.  Their algorithm can be based on the indexes
for jumping to the $a$-labeled children, that is  $\edge/\filtertopath{a}$,
and for jumping to the top-most $a$-labeled descendants, i.e.,
$top_a= (\edge/\filtertopath{\neg a})^*/
\edge/\filtertopath{a}$. 
An \XPath query such as $\texttt{descendant::a}$ can the be rewritten
as the \NRPQ $(top_a)^+$. The evaluation of the query $(top_a)^+$ can then
take advantage of the index $\edge_{top_a}$.
The main difference between both approaches is that ours doesn't try to
produce the answer set in document order, while theirs does so. 
Therefore, binary indexes are sufficient for our 
purpose, while they need to use a ternary index (for relating 
following $a$-labeled nodes $x$ of $y$ below $z$.). Moreover, our algorithm
traverses the same part of the XML document as theirs and will thus be as
efficient while being much simpler in terms of presentation.
Our general graph approach overcomes the main limitations of Maneth
and Nguyen's: it is not bound to trees and is not limited to forward 
navigational XPath but can treat any \NRPQs also with
backward steps.

\begin{toappendix}
  It should be noticed that avoiding indexes of quadratic size may 
be relevant in practice, but more difficult to reach without restrictions.
The index $top_a$,
for instance, may be of the quadratic size, but only for \XML documents 
that do not occur in practice. The choice of appropriate 
indexes raise many interesting research
questions that are out of the scope of the
present paper. It should also be mentioned that 
one may want to represent binary indexes in a more
concise manner, rather than by enumeration of 
node pairs. For instance, for being able to 
jump to $a$-labeled nodes it is sufficient to store all
$a$-labeled nodes, rather than pair of nodes $(x,y)$
such that $y$ is $a$-labeled.
\end{toappendix}

\section{Preliminary Experiments}
\label{sec:expe}
We implemented in OCaml our compiler from PDL to Datalog
and also a compiler from navigational XPath queries
to PDL queries on the graphs of \XML documents. The
edges of these graphs are labeled by $element$, $document$ and
string, and the edges by $first$, $next$, $name$, and attribute
names.

We selected in
in \Figure{XPath} two
typical benchmark \XPath queries from \cite{ManethNguyen10} that can be
applied to the scalable \XML-documents from the XPathMark benchmark:
query $Q01$ composes two child axis, and query $Q05$ two
descendant axis $//listitem//keyword$.  The
translations for these XPath queries to the PDL
queries $pdl.Q01$ and $pdl.Q05$ 
that can also be found there.

Query $pdl.Q01$ is easier for top-down evaluation, since
it does not contain vertically recursive axis, so that its
top-down needed subgraph remains small on the benchmark
documents. $Q05$ is more difficult since using descendant
axis, so that the top-down needed subgraph $pdl.Q05$ is
the whole graph if not using indexes. So we also computed
the indexes $top_{keyword}$ and $top_{listitem}$ for the
descendant axis of $Q05$
and added them as extra edges to the graphs.
Furthermore, the  optimized
query $pdl.Q05.index$ obtained
from $pdl.Q05$ by using the index
edges is given in \Figure{XPath} too.

\begin{figure}[t]
  \begin{minipage}{\textwidth}
  \begin{minipage}{.49\textwidth}
  \begin{tabular}{ll}
  \\\hline
  Q01 & /site/regions 
  \\\hline
  
pdl.  & $\node_{document}?/\edge_{first}/(\edge_{next})^*$/ \\
\ Q01  & $\node_{element}?/[\edge_{name}/\node_{site}?]/$ \\
          & $\edge_{first}/(\edge_{next})^*$/ \\
          & $\node_{element}?/[\edge_{name}/\node_{regions}?]$
     \\    \hline
  \end{tabular}
  \vspace{-.3cm}
\caption{\label{XPath}Two benchmark \XPath queries from \cite{ManethNguyen10},
   their translation to PDL, and the
   indexed PDL queries.}
 \end{minipage}
 \begin{minipage}{.49\textwidth}
   \begin{tabular}{ll}
     \\\hline
Q05 & //listitem//keyword 
\\\hline
pdl.
  &$\node_{document}?/(\edge_{first}/(\edge_{next})^*)^+$/\\
\ Q05   &$\node_{element}?[\edge_{name}/node_{listitem}?]/ $\\
  &$(\edge_{first}/(\edge_{next})^*)^+/        $ \\
  &$\node_{element}?[\edge_{name}/node_{keywords}?]$  
\\\hline
pdl.& $\node_{document}?/(\edge_{top_{listitem}})^+/ $\\
  \  Q05.                  & $(\edge_{top_{keyword}})^+$\\
    \ index &
  \\\hline
   \end{tabular}
\end{minipage}
\end{minipage}
  
   \begin{minipage}{\textwidth}
     \begin{minipage}{.49\textwidth}
\begin{tabular}{llll}
  \\\hline
27KB  &           Q01  &   Q05  &       Q05.index  \\\hline
Saxon &      0.000206  &   indexing & 0.000315  \\
XSB   &        0.001   &   0.006        & 0.001    \\
SWI   &        0.004   &   0.189        & 0.004    \\
  LogicBlox  
      & 0.0045&   0.0054       & 0.0045   \\\hline
\end{tabular}
  \caption{Time in seconds for querying the 27KB XML-document with indexes.}
 \end{minipage}
 \begin{minipage}{.5\textwidth}
\begin{tabular}{llll}
 \\\hline
  100MB &  Q01          &       Q05     &   Q05.index \\\hline
  Saxon &               &    indexing &   0.0016    \\                          
  XSB   &               &   35.857      &   5.029     \\ 
  SWI   &    -          &     -         &      -      \\     
  LogicBlox 
        &0.0124&     -         &   0.0974    \\\hline
\end{tabular}
    \caption{Time in seconds for querying the 100MB
      XML-document with indexes.}
 \end{minipage}
\end{minipage}
  \end{figure}
  The gold standard for the evaluation of XPath queries is obtained
  by using the Saxon XSLT evaluator. In order to measure the time we run the
same XPath query 100 times in the same XSLT program with Saxon  10.5,
substract the time needed to load and index the XML document
and divide by 100. It turns out, that Saxon has the best
performance in all our tests, confirming our conjecture
that it performs jumping evaluation with indexing for
descendant axis.

We then implemented and
tested our jumping algorithm based on existing
top-down Datalog evaluators. We started with
OCaml's Datalog 0.6, but had to notice that the
top-down evaluator did not always produce the correct results.
  We then experimented with the Prolog engines XSB 4.0 and SWI 8.4.1.1. 
  On a small XML-document of 27KB, both engines perform
  decently, even though not as quick as Saxon. On Q01, they
  are one order of magnitude slower. The same holds for Q05 but
  only when using indexing.

  We then considered a much bigger XML-document of 100MB.
  With this size we had to give up with SWI. XSB in contrast
  could read the graph of the XML document, but needed
  more than 30 minutes. Once the graph was read, it could answer the query
  $pdl.Q05$ without indexing in 35 seconds. With
  indexing the time for answering $pdl.Q05.index$ went  down
  to $5$ seconds. Saxon, in contrast, can load the graph in 15 seconds
  and answer query $Q05$ in $1.6$ milliseconds. So for answering the
  query $Q05$,  Saxon showed $4$ orders of magnitude
  more efficient than XSB.

  We finally investigated the LogicBlox system \cite{LogicBlox-short}, a
  more  recent deductive database system which implements the
  language LogiQL extending on Datalog. With version 4.38 of
  LogicBlox we could read the graph of 100 MB in 19 seconds (rather
  then in more than 30 minutes as with XSB).
  LogicQL is a typed language implying some minor
  syntactic differences to standard datalog. Finally, LogicBlox
  has a transaction level, that permits to interact with graphs
  dynamically, so that it can be queried many times without
   being reloaded.
   The earlier versions of LogicBlox supported bottom-up 
   evaluation only. But since recently, top-down evaluation can
   be chosen by adding On-Demand annotations for
   all extensional predicates. When doing so, we could
   answer the query $pdl.Q05.index$  in 97.4 milliseconds on
   the 100MB document. This is 2 orders of magnitudes better
   than  with XSB!   Nevertheless it is still by a factor of 75 slower than
   with Saxon. Figure 4 of \cite{ManethNguyen10} reports ~65
   milliseconds for Q05 with optimal jumping, but
   on a slightly larger 116MB document.   So the question is how the efficiency of
   our implementation could be increased further: with 
   better indexes, early completion during Datalog evaluation,
   or by using special features of \XPath queries?

\section{Conclusion and Future Work}
\label{sec:future}
The definition of the top-down needed subgraph allows us to prove that
our algorithm for answering negation-free \NRPQs visits only the 
interesting part of the graph. We believe that the restriction to 
negation-freeness can be relieved by compiling to stratified
datalog.
The new notion of top-down needed subgraphs may also allow the 
design of algorithms that transform \NRPQs into equivalent ones that have a
smaller top-down needed subgraph, for instance by inverting the path, 
or starting with some filter. 
In particular, the $\goto$ instructions permit the algorithm to jump
directly to nodes with \emph{rare} properties in the graph first and then
compute the queries more efficiently.
Another line of improvement would be to stop the evaluation of a filter when it
has been proven correct. This effect may only be obtained if we use
a datalog top-down evaluator that follows the \emph{early completion}
strategy, i.e. stops whenever a ground predicate (such as filter queries in our case)
is proven true.

\ignore{
\begin{toappendix}
 Another line of improvement would be to stop the evaluation of filters when it
has been proven correct.
 This effect may only be obtained if we use
a datalog top-down evaluator that follows the \emph{early completion}
strategy, i.e. stops whenever a ground predicate (such as filter queries in our case)
is proven true. But this strategy does not survive magic-set rewriting
of Datalog programs, in order to mimic top-down evaluation in a 
bottom-up manner.
Moreover, \emph{early completion} does not allow us to define clearly a notion
of needed nodes in a graph for a given query. A way out of this problem is to
implement directly in datalog what it means to traverse a set of nodes
sequentially. For this, we need to assume that outgoing edges in graphs are
ordered. This local order extends to a total order on paths starting at a given
node by using a lexicographic order. Then we may implement a depth-first
left-to-right traversal of the graph following this lexicographic
order.
\end{toappendix}}
%



\bibliographystyle{eptcs}
\bibliography{mostrare.bib,tom.bib,new.bib}


\end{document}